\documentstyle[12pt]{article}
\title{Effects of the final-state interaction
in ($\gamma$,pn) and ($\gamma$,pp) processes}
\author{J. Ryckebusch \thanks{Postdoc research associate NFWO},
M. Vanderhaeghen \thanks{Research fellow NFWO},
L. Machenil \thanks{Research fellow IWONL} \\
and M. Waroquier \thanks{Research director NFWO} \\[2.cm]
Laboratory for Nuclear Physics and Laboratory for Theoretical
Physics \\ Proeftuinstraat 86 \\ B-9000 Gent, Belgium}
\begin{document}
\maketitle
\begin{abstract}
A model is presented to describe electromagnetically induced two-nucleon
emission processes in a shell-model picture.  Distortions in the
outgoing nucleon waves are accounted for by performing a partial-wave
expansion in a real mean-field potential.  The antisymmetry condition
for the A-body wavefunctions is shown to be naturally preserved.
  The model is used to calculate
($\gamma$,pn)  and ($\gamma$,pp) cross sections off the target nuclei
$^{16}$O and $^{12}$C for
photon energies ranging from  50 MeV up to the $\bigtriangleup$(1232)
 isobar threshold.
Effects due to the pionic currents and intermediate $\bigtriangleup$ creation
are implemented.  The impact of the distortions due to the interaction
of the outgoing nucleon waves with the (A-2) core is
examined.  Hadronic form factors are introduced to regularize the $\pi$NN
vertices and the sensitivity of the cross section to the pion cut-off mass
is examined.  The relative contribution of the ($\gamma$,pp) and the
($\gamma$,pn) channel to the total photoabsorption strength is discussed.
Further, the photon energy dependence of the ($\gamma$,pp)/($\gamma$,pn)
ratio is investigated.
\end{abstract}
\addtolength{\baselineskip} {0.5 \baselineskip}
\section{Introduction}
Following the suggestion that the ($\gamma$,pn) reaction
might be a useful tool for the exploration of short-range
correlations (SRC) in finite nuclei \cite{Got58}, the photoexcitation of
correlated proton-neutron pairs has been studied
extensively.   The exploratory theoretical investigations
date back to the sixties and the seventies.
Rather representative for the early ($\gamma$,pn) investigations are the
$^{16}$O($\gamma$,pn)
calculations by W. Weise {\em et al.}.   In their approach
the photon was assumed to
couple exclusively to the one-body convective current.  At the
same time a proton-neutron correlation function of the Jastrow
type ($1-j_0(q\mid{\bf r}_1-{\bf r}_2\mid)$)
was introduced in the shell-model wave functions.
In this way a two-body photoabsorption process
was generated which can naturally explain the emission of a nucleon
pair.   The calculations of ref. \cite{Wei70} along this vein
clearly illustrate  that the
absolute magnitude of the cross sections is very sensitive to the
value of the correlation parameter q.  With a reasonable value of the
correlation parameter q the calculations were found to account for
the absolute magnitude of the measured ($\gamma$,pn) strength. In
most of the ($\gamma$,pn) calculations of the seventies the
meson-related contribution to the pn strength was assumed to be
unimportant \cite{Bra71}. \\
At present the physical context of ($\gamma$,pn) reactions is quite
different.  Recently it has been recognized that photoabsorption on
two-body currents of mesonic nature is responsible for a large
fraction of the measured ($\gamma$,pn) strength
\cite{Wak83,Bof91,Boa89,Ryc92}.   Therefore, the hope
to observe direct indications of short-range correlations in nuclei
has somewhat shifted from the dominant ($\gamma$,pn)
to the weaker ($\gamma$,pp) channel which cannot
be fed in a direct two-nucleon knockout mechanism
following photoabsorption on a
pionic current.  For that reason the pp channel is suggested to be more
sensitive to the SRC. Amongst the reaction mechanisms which are most
likely to
contribute to the ($\gamma$,pp) strength we mention : short-range
correlations of the Jastrow type, intermediate delta excitation and
the coupling with the strong ($\gamma$,pn) channel.  The latter
involves a multi-step process consisting of an initial $\gamma + p + n
\longrightarrow n + p $ and a rescattering process of the (n,p) type.
A similar type of channel-to-channel coupling was suggested
\cite{Gar81,Ryc88}
to be essential in order to explain the observed strength in
 the photoninduced one-nucleon emission channels
($\gamma$,p) and ($\gamma$,n) where a considerable fraction of the
($\gamma$,n) strength has been ascribed to initial photoabsorption on
a proton and a rescattering process of the (p,n) type.
The predicted coupling \cite{Bau90}
between the strong and the weaker channels suggests
that a full understanding of the two-nucleon emission processes will
only be reached after an in-depth investigation of
both the ($\gamma$,pn) and ($\gamma$,pp) channels.

The advent of high-duty cycle tagged photon facilities has revived
the field of ($\gamma$,NN) reactions as a probe for
investigating nucleon-nucleon correlations in nuclei.  On the
theoretical side, the quantitative contribution of the
meson exchange currents to the ($\gamma$,pn) cross sections was first
assessed by Gari and Hebach with the aid of Siegert's theorem \cite{Gar81}.
In the Fermi-gas ($\gamma$,pn) studies by Wakamatsu and Matsumoto
a more fundamental approach for the coupling of the photon with the
mesons was followed \cite{Wak83}.  Their current operators were derived
through minimal substitution in the one-pion exchange potential. More
recently, Boato and Gianninni calculated the contribution of the
pion-exchange currents to photoinduced two-nucleon emission on finite
nuclei in a factorized approach \cite{Boa89}.
It should be stressed that several
assumptions have to be made before the ($\gamma$,NN) cross section
transforms in a factorized form.  First, all interaction effects between
the outgoing nucleon pair and the core together with the mutual
interactions of the two escaping nucleons have to be neglected.
Furthermore, the reaction is assumed to take place at sufficiently small
internucleon distances so that only nucleon pairs moving in relative S
waves are believed to take part in the interaction with the external photon
field (quasi-deuteron approximation).   Lastly, the radial dependence
of the relative S waves must be discarded in order to achieve full
factorization
(zero-range approximation).    The interaction between the
outgoing nucleon pair and the residual A-2 core has been recently
investigated by Boffi and Giannini in an optical-potential model approach
\cite{Bof91}.  In these calculations the cross section was assumed to
retain its factorized form even when distortions are accounted for.

In a previous publication \cite{Ryc92} we have shown that even when
neglecting all distortion effects important deviations between the
results obtained with the factorized and the full (unfactorized) cross
section are observed.    In this paper we aim at developing an
unfactorized model for two-nucleon emission reactions
which remains numerically
tractable and in which the interaction effects between the escaping
nucleons and the core can be calculated.

The plan of this paper is as follows.  In sect. 2
we present a shell-model approach to two-nucleon
emission processes in which we account for the distortions in the
outgoing particle waves related to the interaction with the residual A-2
core. The method involves a partial wave expansion of the emitted nucleon
waves and does naturally preserve the antisymmetrization between the
escaping particles and the core.  In sect. 3 the model is applied to
the calculation of
$^{12}$C($\gamma$,pp), $^{16}$O($\gamma$,pp), $^{12}$C($\gamma$,pn) and
$^{16}$O($\gamma$,pn)
cross sections in the
photon energy range below the $\bigtriangleup$ resonance.  In the
process of calculating the cross sections we account for
photoabsorption on the pionic currents (Seagull and pion-in-flight
diagram) and intermediate $\bigtriangleup$ creation followed by the
emission of two nucleons.  We
investigate the impact of the distortions on the cross sections and
the role of the different reaction mechanisms.  A considerable fraction
of sect. 3 is devoted to the study of the contribution
of the ($\gamma$,pn) and ($\gamma$,pp) channel to the total
photoabsorption strength.
Finally, our
conclusions will be given in sect. 4.

\section{The formalism}

\subsection{Derivation of the ($\gamma$,NN) cross section}

In this paper we will be dealing with photoinduced two-nucleon
emission processes, viz reactions of the type
\begin{equation}
\gamma + A \longrightarrow N_a + N_b + (A-2) \; ,
\end{equation}
where N$_a$ and N$_b$ refer to the two escaping nucleons following the
absorption of a photon with energy $E_{\gamma}$ by the target nucleus
A.  The cross section for the considered process can be shown to take
on the following form in the lab frame
\begin{eqnarray}
\frac{d^5 \sigma^{lab}} {d\Omega_a d\Omega_b dk_b} = \frac {1} {(2
\pi)^5} \sum _{f}
\frac{\left| m_F^f \right|^2 k_a^2 E_{A-2} E_a k_b^2}
{2E_{\gamma}\left[k_a(E_a+E_{A-2})-E_a
(q_{\gamma}cos\theta_a-k_bcos\theta_r)\right]}
\;,
\label{five}
\end{eqnarray}
where ${\bf k}_a ({\bf k}_b)$ is the lab momentum of the escaping
nucleon N$_a$ (N$_b$), $\theta_r$ is the angle between ${\bf k}_a$
and ${\bf
k}_b$ and $E_{A-2}$ is the total energy of the residual nucleus in the
lab frame.  We have chosen the z axis along the direction of ${\bf
q_{\gamma}}$.  All the nuclear dynamics information is contained in
the Feynman amplitude $m_F^f$ which is defined according to :

\begin{equation}
m_F^f=\int d{\bf r} e^{i{\bf q}_{\gamma} \cdot {\bf r}} < \Psi_f \mid
{\bf J} ({\bf r}) \cdot \mbox{\boldmath$\epsilon$}_{\lambda} \mid \Psi_i > \; ,
\label{feynman}
\end{equation}
where ${\bf J}$ is the nuclear current operator and \mbox{\boldmath$
\epsilon$}$_{\lambda}$ the polarization vector of the photon.  The choice of
the
current operator is related to the model assumptions which are made
with respect to the photoabsorption mechanism.  This will be the
subject of discussion in sect. 2.3. In the forthcoming section
we will address the problem of how to construct appropriate A-body
wavefunctions $\mid \Psi_i>$ and $\mid \Psi_f>$
which can be used to calculate
the angular cross sections for electromagnetically induced
two-nucleon emission processes.  For the time being we will not
introduce any SRC corrections to the shell-model wave functions.
Accordingly, all the nuclear wave functions of this work are Slater
determinants.

\subsection{Shell-model wave functions with two particles in the
continuum}

Despite the fact that the basic electromagnetic interaction of (virtual)
photons with the nucleus is relatively well understood,
the analysis of coincidence experiments of the type ($\gamma$,N) and
(e,e$'$N) on finite nuclei
is often hampered by the strong interactions between the
struck nucleon and the medium in which it is embedded.  Only under
exceptional circumstances it was found that the (A-1) nucleons will
act as spectators.  In the most general case a complex variety of
strong interactions between the escaping nucleon and the residual
core will affect the cross section.  These processes are
commonly referred to as the final state interaction (FSI).  With
respect to the FSI two types of effects can be discriminated.  First
we have to realize that the escaping nucleon cannot be looked upon as
a free particle.  Its wave function will somehow reflect the
distortions it undergoes in its way out of the nucleus.  Secondly, a
whole class of multi-step processes can finally lead to an escaping
nucleon in a particular channel.  In fig. 1 we have sketched some
diagrams which are considered to represent the main contributions
to the FSI in a one nucleon knockout process.
In drawing these diagrams we have restricted ourselves
to the case in which the photon is absorbed on one nucleon.
All nucleon lines in the
diagrams of fig. 1 have to be associated with mean-field single-particle wave
functions.
Diagram 1(a) refers to the most simple case.  The struck
nucleon is excited in a continuum state of the real mean-field potential
in which also the bound-state single-particle wave functions are
generated. The processes of diagram 1(b) will generally lead
to a reduction of the cross section and can be modeled
for through generating the distorted wave for the escaping nucleon in
a complex optical potential.  The diagrams of fig. 1(c) are more
difficult to implement since they involve a coupled channel
calculation of the RPA or continuum shell-model type.
 Depending on the kinematical conditions
the diagrams of fig.1(c)
will lead to a reduction or increase of the strength in a
particular channel.
Throughout the years a combined effort of several groups has resulted
in a reasonable understanding of the FSI effects for the case of one
escaping nucleon.

\begin{figure}[tbh]
\vspace{6.5 cm}
\caption{Different classes of diagrams for the FSI in
electromagnetically induced one-nucleon emission processes.  All nucleon
lines refer to eigenfunctions of a mean-field potential.
Photoabsorption on one nucleon is assumed.}
\end{figure}

The situation in which two nucleons are ejected in coincidence from a
target nucleus represents a complicated three-body problem
and model
assumptions regarding the FSI could be expected to be
indispensable in order to keep the calculations feasible.  An equivalent set
of diagrams for those depicted in fig. 1 for the one-nucleon emission
case are shown in fig. 2.  For the case of two-nucleon emission we
have restricted ourselves to two-body photoabsorption.  Diagram 2(a)
is the equivalent of 1(a) for the one-body case.  The processes of
fig. 2(b) involve an interaction of each of the escaping nucleons
with the core and can be accounted for by calculating both
distorted waves in an optical potential.  The recent calculations by
the Pavia group \cite{Giu92} have been performed along this line.
In these calculations, the sole impact of the imaginary part
of the potential was found to
be an overall reduction of the cross section.
The processes sketched in fig. 2(c) involve an interaction between
the two struck nucleons, whereas for diagram 2(d) both detected
nucleons are assumed to have undergone RPA-like rescattering effects.
To our knowledge, no model is available which has estimated the
general effect of rescattering effects in ($\gamma$,NN) or (e,e$'$NN)
processes.

In what follows we will restrict ourselves to the
diagrams of the type sketched in fig. 2(a).  It involves direct
two-nucleon knockout following photoabsorption on a two-body current.
The bound and the continuum states will be generated in the same
mean-field potential.  In this way we preserve the orthogonality
between the nuclear states.

\begin{figure}[tbh]
\vspace{10. cm}
\caption{Different classes of diagrams for the FSI in
electromagneticaly induced two-nucleon emission. Two-body photoabsorption
is assumed.}
\end{figure}

Dealing with two-nucleon emission processes in a shell-model
picture, we have to
construct eigenstates $\mid \Psi_f>$ of the many-body Hamiltonian, that
asymptotically for two arbitrary coordinates ${\bf r}_1$ and ${\bf r}_2$
tending to infinity, behave like

\begin{eqnarray}
\left<{\bf r}_1 \mbox{\boldmath$\sigma$}_1, {\bf r}_2
\mbox{\boldmath$\sigma$}_2
    \mid \Psi_f
\right> & \stackrel {r_1,r_2 \gg r_A} {\longrightarrow} & \frac {1}
{\sqrt{A(A-1)}} {\cal A}_{2(A-2)} \nonumber \\
 & & \left[ \chi_{\frac{1}{2}m_s} (\mbox{\boldmath$\sigma$}_1)
\left(e^{i{\bf k}_{a} \cdot
{\bf r}_1} + f_{k_{a}}(\theta_a) \frac {e^{ik_ar_1}} {r_1} \right)
\right. \nonumber
\\ & & \times \left.
\chi_{\frac{1}{2}m_{s'}} (\mbox{\boldmath$\sigma$}_2) \left(e^{i{\bf k}_b {\bf
\
   cdot
r}_2} + f_{k_{b}}(\theta_b) \frac {e^{ik_br_2}} {r_2} \right) \mid
(hh')^{-1} J_R M_R > \right] \;,
\end{eqnarray}
where ${\cal A}_{2(A-2)}$ antisymmetrizes the wave functions for the
outgoing nucleons with respect to themselves and with respect to the
residual (A-2) system.    In the above wave function the residual (A-2)
nucleus is described by the two-hole ({\em 2h}) state $\mid (hh')^{-1}
J_R M_R >$ which is
defined according to

\begin{eqnarray}
\mid (hh')^{-1}J_R M_R > = \sum_{m_hm_{h'}} \frac{1}{\sqrt{1+\delta_{hh'}}}
<j_hm_hj_{h'}m_{h'} \mid J_R M_R > \nonumber \\
\times (-1)^{j_h+m_h+j_{h'}+m_{h'}} c_{h-m_{h}}
c_{h'-m_{h'}} \mid \Phi_0 >\;.
\end{eqnarray}
The {\em 2h} state $\mid J_R M_R>$ is a fully antisymmetrized and normalized
(A-2) wave function and $\mid \Phi_0>$ is the uncorrelated
ground-state wave function
of the target nucleus.   The wave function of eq. (4) refers to the situation
where two nucleons are escaping from the system with momentum ${\bf k}_a
({\bf k}_b)$ and spin projection $ m_{s} ( m_{s'})$.
In case of an induced one-particle emission process in which a nucleon
is ejected from the target nucleus with a momentum {\bf k}  and in which
the residual nucleus is residing in a pure hole state {\em h}, an appropriate
asymptotic wavefunction
\begin{equation}
\frac{1}{\sqrt A} {\cal A} \left[
\chi_{\frac{1}{2}m_s} (\mbox{\boldmath$\sigma$}_1) \left(e^{i{\bf k \cdot
r_1}} + f_k(\theta) \frac {e^{ikr_1}} {r_1} \right) (-1)^{j_h+m_h} c_{h-m_h}
\mid \Phi_0> \right]
\end{equation}
is normally obtained by performing a multipole expansion in terms of
elementary particle-hole ($ph$) excitations \cite{Ryc88,Mah} :
\begin{eqnarray}
\mid \Psi_f > & = & \sum_{lm_ljm} \sum_{JM} 4 \pi i^l
\sqrt{\frac {\pi} {2 \mu k}} <j_hm_hjm \mid JM> \nonumber \\
              & & \times <lm_l\frac{1}{2}m_s\mid
jm> e^{i(\delta_l+\sigma_l)}Y_{lm_l}^{*}(\Omega_k)
\mid p(\epsilon) h^{-1}(JM)>\;,
\end{eqnarray}
where $\epsilon \equiv k^2/(2 \mu)$,
$\mu$ is the reduced mass of the outgoing nucleon, $\delta _l$ is the
central phase shift, $\sigma _l$ is the Coulomb phase shift and $\Omega _k$
determines the solid angle of the momentum ${\bf k}$ of the escaping
nucleon.
The above expression has been derived under the following normalization
conventions
for the continuum eigenstates $\varphi_{lj}$
of the real mean-field potential~:
\begin{equation}
\varphi_{lj}(r,E) \stackrel{r \gg R} {\longrightarrow}
\sqrt{\frac{2\mu} { \pi k}} \frac {sin(kr-\eta ln(2kr) -
\frac{\pi l}{2} + \delta _l + \sigma _l)} {r}\;.
\end{equation}

As a natural extension of the above partial-wave expansion in terms
of $ph$ excitations for one-particle emission processes, we suggest
the following partial-wave expansion in terms of $2h2p$ states for
the case of two-nucleon emission processes :
\begin{eqnarray}
\mid \Psi_f > & = & \sum_{lm_ljm}\sum_{l'm_{l'}j'm'} \sum_{JMJ_1M_1}
(4 \pi)^2 i^{l+l'}
\frac {\pi} {2 \mu \sqrt{k_a k_b}}
e^{i(\delta_l+\sigma_l+\delta_{l'}+\sigma_{l'
   })}
Y_{lm_l}^{*}(\Omega_{k_{a}}) Y_{l'm_{l'}}^{*}(\Omega_{k_{b}})  \nonumber \\
              & & \times <lm_l\frac{1}{2}m_s\mid
jm>  <l'm_{l'}\frac{1}{2}m_{s'}\mid j'm'>
<J_R M_R J_1 M_1 \mid J M > \nonumber \\
& & \times < j m j' m' \mid J_1 M_1 >
\mid (hh')^{-1} J_R ; (p(\epsilon _a)p'(\epsilon _b)) J_1 ; JM> \;,
\label{wave}
\end{eqnarray}
where $\epsilon_a \equiv k_a^2/(2 \mu)$, $\epsilon_b \equiv k_b^2/(2
\mu)$ and
the ${\em 2h2p}$ state $\mid (hh')^{-1} J_R , (pp') J_1 ; JM>$
is defined according to :
\begin{eqnarray}
\sum_{mm'}\sum_{M_1M_R} <jmj'm' \mid J_1 M_1> <J_R M_R J_1 M_1 \mid J
M > c_{ljm}^{\dagger}c_{l'j'm'}^{\dagger} \mid (hh')^{-1} ; J_R M_R>\;.
\end{eqnarray}
Since the wave function of eq. (\ref{wave}) is a linear combination of ${\em
2h2p}$ Slater determinants it obeys the antisymmetry condition.    It
can be easily shown that the wave function of eq. (\ref{wave}) has the
required asymptotic behaviour with two escaping particles and a
residual (A-2) system remaining in the state $\mid J_R M_R >$.

\subsection{Absorption mechanisms and transition matrix elements}

Having established the proper nuclear wave functions we remain with
determining the dominant photoabsorption mechanism.  In this work we
have restricted ourselves to the diagrams with one pionic line as
indicated in fig. 3.  Diagrams (a) and (b) correspond to the coupling
of the electromagnetic field with the pionic current and are commonly
referred to as the seagull ($\pi sea$) and the pion-in-flight ($\pi
fli$) diagrams.  Next, diagrams (c) are related to intermediate
$\bigtriangleup$-isobar creation.  In this paper we have employed a
pseudovector $\pi$NN coupling.   Given the charge-exchange nature
of diagrams 3(a) and 3(b) they will only contribute to the
($\gamma$,pn) channel.  Of all of the diagrams of fig. 3 the
pion-in-flight term is the most difficult one to calculate, since it
does involve two pion propagators.

\begin{figure}[tbh]
\vspace{10. cm}
\caption{Absorption mechanisms which are accounted for in the
present calculations.}
\end{figure}

In r-space the currents corresponding to diagrams (a) and (b) read
\begin{eqnarray}
{\bf J}^{(\pi sea)}({\bf r},{\bf r}_1,{\bf r}_2) & = & e \left( \frac {f_{\pi
NN}} {m_{\pi}} \right)^2 \left(\mbox{\boldmath$\tau$}_1 {\bf \times}
\mbox{\bold
   math$\tau$}_2
\right)_z \left\{\delta({\bf r} - {\bf r}_2) \mbox{\boldmath$\sigma$}_2
(\mbox{\boldmath$\sigma$}_1 {\bf \cdot} \mbox{\boldmath$\nabla$}_1) \right.
\non
   umber \\
& & \left. - \delta({\bf r} - {\bf r}_1) \mbox{\boldmath$\sigma$}_1
(\mbox{\boldmath$\sigma$}_2 {\bf \cdot} \mbox{\boldmath$\nabla$}_2) \right\}
\frac {e^{-m_{\pi} \mid {\bf r}_1 - {\bf r}_2 \mid}}
      {4 \pi \mid {\bf r}_1 - {\bf r}_2 \mid} \;, \\
{\bf J}^{(\pi fli)}({\bf r},{\bf r}_1,{\bf r}_2)
& = & e \left( \frac {f_{\pi
NN}} {m_{\pi}} \right)^2 \left(\mbox{\boldmath$\tau$}_1 {\bf \times}
\mbox{\boldmath$\tau$}_2
\right)_z \left\{ \mbox{\boldmath$\sigma$}_1 {\bf \cdot}
\mbox{\boldmath$\nabla$}_1
\mbox{\boldmath$\sigma$}_2 {\bf \cdot} \mbox{\boldmath$\nabla$}_2
(\mbox{\boldmath$\nabla$}_2 -\mbox{\boldmath$\nabla$}_1)
\right\} \nonumber \\
& & \times
\frac {e^{-m_{\pi} \mid {\bf r} - {\bf r}_1 \mid}}
      {4 \pi \mid {\bf r} - {\bf r}_1 \mid}
\frac {e^{-m_{\pi} \mid {\bf r} - {\bf r}_2 \mid}}
      {4 \pi \mid {\bf r} - {\bf r}_2 \mid} \;.
\end{eqnarray}
In the static limit,
the diagrams of fig. 3(c) give rise to the following current :
\begin{eqnarray}
{\bf J}^{(\pi \bigtriangleup)}({\bf r},{\bf r}_1,{\bf r}_2) & = &
\frac {2 f_{\gamma N \bigtriangleup} f_{\pi N \bigtriangleup} f_{\pi
NN}} {9 m_{\pi}^3 (M_{\bigtriangleup} - M_N)}
\left\{ \left[ \left(\mbox{\boldmath$\tau$}_1 {\bf \times}
\mbox{\boldmath$\tau$}_2
\right)_z \mbox{\boldmath$\sigma$}_2 {\bf \cdot} \mbox{\boldmath$\nabla$}_2
(\mbox{\boldmath$\sigma$}_1 \times
\mbox{\boldmath$\nabla$}_2) \times (\mbox{\boldmath$\nabla$}_1 +
\mbox{\boldmath$\nabla$}_2) \right. \right. \nonumber \\ +
& & \left. \left . 4 (\mbox{\boldmath$\tau$}_2)_z \mbox{\boldmath$\sigma$}_2
{\bf \cdot} \mbox{\boldmath$\nabla$}_2
(\mbox{\boldmath$\nabla$}_1 {\bf \times} \mbox{\boldmath$\nabla$}_2)
 \delta({\bf r} - {\bf r}_1)
 \right]
+ 1 \longleftrightarrow 2 \right\}\frac {e^{-m_{\pi}
\mid {\bf r}_2 - {\bf r}_1 \mid}}
      {4 \pi \mid {\bf r}_2 - {\bf r}_1 \mid}\;.
\label{eq.del}
\end{eqnarray}
At higher photon energies the above ${\bf J}^{(\pi \bigtriangleup)}$
current has
to be corrected for the finite lifetime of the $\bigtriangleup$(1232),
whereas also $\rho$ decay of the resonance could be expected to
start playing a role.  For the present purposes we have utilized the
static current of eq. (\ref{eq.del}).  A more elaborate treatment of the
${\bf J}^{(\pi \bigtriangleup)}$ current falls beyond the scope of the
present work and will be presented elsewhere \cite{Mac93}.
In order to regularize the above currents at short internucleon
distances, hadronic form factors have to be introduced at each
$\pi NN$ and $\pi N \bigtriangleup$ vertex.  As is commonly done we have
adopted a monopole parametrization for the form factor.  After
introducing the same monopole form factor at each $\pi NN$ and $\pi N
\bigtriangleup$ vertex, the pion propagators
\begin{eqnarray}
\frac {e^{-m_{\pi} \mid {\bf r}_2 - {\bf r}_1 \mid}}
{4 \pi \mid {\bf r}_2 - {\bf r}_1 \mid} \equiv \frac{1}{(2\pi)^3}
\int d{\bf p} \frac{e^{i {\bf p \cdot} ( {\bf r}_1 - {\bf
r}_2)}} {p^2+m_{\pi}^2} \; ,
\end{eqnarray}
in the above currents will be replaced by :
\begin{equation}
\frac{1}{(2\pi)^3}
\int d{\bf p} \frac{e^{i {\bf p \cdot} ( {\bf r}_1 - {\bf
r}_2)}}{p^2+m_{\pi}^2} \left(\frac{\Lambda_{\pi}^2 - m_{\pi}^2}
{\Lambda_{\pi}^2 +p^2} \right)^{\beta} \; ,
\end{equation}
where $\Lambda_{\pi}$ is the so-called pion cutoff mass and $\beta$=2
for {\bf J}$^{(\pi sea)}$ and {\bf J}$^{(\pi \bigtriangleup)}$, whereas
$\beta$=1 for {\bf J}$^{(\pi fli)}$.  Recent
parametrizations of the Bonn potential lead to $\Lambda_{\pi}$=1200
MeV \cite{Bonn}.  On the other hand, triton binding energy and
Gamow-Teller studies seem to prefer $\Lambda_{\pi}$=810
MeV \cite{Sas92}. A similar value was found to produce the best
results by Wakamatsu and
Matsumoto in their $^{9}$Be($\gamma$,pn) and
$^{9}$Be($\gamma$,p$\pi^-$) investigations in a Fermi-gas model
\cite{Wak83}.
\\

Having adopted model assumptions for the nuclear wave functions and
the coupling of the electromagnetic photon field to the nucleus, we
are now in the position to calculate the Feynman amplitude of eq.
(\ref{feynman}).
Since we have performed an angular momentum expansion for the final
states we are forced to apply a similar procedure to the current
operator.  This can be easily achieved by performing a multipole
decomposition
in terms of the electric and magnetic transition
operator.  The Feynman amplitude of eq. (\ref{feynman})
can then be written as :
\begin{equation}
m_F^f= - \sqrt{(2 \pi)} \sum _{J \geq 1} i^J \hat{J} < \Psi_f \mid
T^{el}_{J \lambda} (q_{\gamma}) + \lambda T^{mag}_{J \lambda}
(q_{\gamma}) \mid \ \Psi_i> \;,
\end{equation}
where the electric and magnetic transition operator are defined
according to the conventions of ref. \cite{For66} and $\hat{J} \equiv
\sqrt{2J+1}$.  Inserting the wave
function of eq. (\ref{wave}) in the above expression, the Feynman
amplitude for the process in which the absorption of a photon
with polarization $\lambda$ is followed by the emission of a nucleon
pair ({\bf k}$_a$,m$_{s}$;{\bf k}$_b$,m$_{s'}$)
becomes~:
\begin{eqnarray}
m_F^f & = & - \sqrt{2 \pi} \sum _{J \geq 1} i^J \hat{J}
\sum_{lm_ljm}\sum_{l'm_{l'}j'm'} \sum_{J_1M_1}
(4 \pi)^2 (-i)^{l+l'}
\frac {\pi} {2 \mu \sqrt{k_a k_b}} e^{-i(\delta_l+\sigma_l
+\delta_{l'}+\sigma_{l'})}
  \nonumber \\
& & \times Y_{lm_l}(\Omega_{k_{a}}) Y_{l'm_{l'}}(\Omega_{k_{b}})
<lm_l\frac{1}{2}m_s\mid
jm>  <l'm_{l'}\frac{1}{2}m_{s'}\mid j'm'> \nonumber \\
& & \times < j m j' m' \mid J_1 M_1 >
\frac{(-1)^{J_R-M_R+1}}{\hat{J_1}}
<J_R -M_R J \lambda \mid J_1 M_1> \nonumber \\
& & \times <p(\epsilon _a) p'(\epsilon _b);J_1
\| T^{el}_{J} (q_{\gamma}) + \lambda T^{mag}_{J}
(q_{\gamma}) \|hh' ; J_R >_{as} \;,
\label{feynb}
\end{eqnarray}
where the antisymmetrized two-body matrix elements are defined according
to :
\begin{eqnarray}
<pp';J_1 \| T_{J} (q_{\gamma}) \|hh' ; J_R >_{as} & \equiv &
<pp';J_1 \| T_{J} (q_{\gamma}) \|hh' ; J_R> \nonumber \\
& & - (-1)^{j_h+j_{h'}+J_R}
<pp';J_1 \| T_{J} (q_{\gamma}) \|h'h ; J_R>\;.
\end{eqnarray}
For the reduced matrix elements we adopt the conventions of ref.
\cite{talmi}.
Referring to eq. (\ref{feynb}),
the calculation of the ($\gamma$,NN) cross sections has been
reduced to determining antisymmetrized two-body matrix elements of the type
$<pp';J_1 \| T^{el}_{J} (q_{\gamma}) \|hh' ; J_R >_{as}$ and
$<pp';J_1 \| T^{mag}_{J}
(q_{\gamma}) \|hh' ; J_R >_{as}$.
  These matrix elements are a function of the adopted photoabsorption
mechanism and have
been summarized in Appendix A for the diagrams of fig. 3.

The contribution from the ($\gamma$,N$_{a}$N$_{b}$) channel to the total
photoabsorption strength $\sigma$($\gamma$,N$_{a}$N$_{b}$) can be
calculated by integrating the fivefold differential cross section over
the respective solid angles and nucleon momentum {\bf k}$_b$ :
\begin{equation}
\sigma(\gamma,N_{a}N_{b}) = \int d \Omega_a d \Omega _b dk_b
\frac{d^5 \sigma^{lab}} {d\Omega_a d\Omega_b dk_b} \;.
\label{inte}
\end{equation}
After averaging over the intial photon polarization and summing over the
spin polarizations of the escaping nucleons it can be shown that upon
substituting the Feynman amplitude of eq. (\ref{feynb}) in the fivefold
differential cross section (\ref{five}) and integrating over both solid
angles the following expression is obtained :
\begin{eqnarray}
\sigma(\gamma,N_{a}N_{b}) & = & \int  dk_b \frac{E_a E_{A-2} k_b \pi}
{E_{\gamma} E_A \mu^2} \sum_{J\geq 1} \sum_{J_1 lj l'j'}
\left\{
\left| <pp';J_1 \| T^{el}_{J} (q_{\gamma}) \|hh' ; J_R >_{as} \right|^2
\right. \nonumber \\
& & +  \left. \left| <pp';J_1 \| T^{mag}_{J}
(q_{\gamma}) \|hh' ; J_R >_{as} \right|^2 \right\} \;.
\label{eq:inte}
\end{eqnarray}
In deriving this expression we have neglected
the effect of nuclear recoil.


\section{Results}
In this section we report on the results of the numerical
calculations for the $^{16}$O($\gamma$,pn),
$^{12}$C($\gamma$,pn),
$^{16}$O($\gamma$,pp) and $^{12}$C($\gamma$,pp)
cross sections which we have
performed in the theoretical framework discussed above.  The
presented calculations primarily aim at exploring the main sensitivities
in the calculation of ($\gamma$,pp) and ($\gamma$,pn) cross sections.
Given that we have assumed that the photon couples exclusively to the
lightest meson and that apart from the $\bigtriangleup$ no other
intermediate nucleon resonances are created, the scope of the
present calculations has to be restricted to the photon energy range
below 300 MeV.  On the other hand,
given the relatively small momentum transfer involved in real
photon reactions, the coupling of the photon to the one-pion exchange
part of the NN interaction in combination with intermediate
$\bigtriangleup$ creation could be expected to set the main trends of
the ($\gamma$,pp) and ($\gamma$,pn) cross sections in the considered
energy range.  However, only a detailed confrontation
of the calculated angular
cross sections with the data will rule out whether other
effects, like SRC effects e.g.,
play a major role in the photon energy range below the
$\bigtriangleup$ production threshold. At present, none of the data
have been presented in a model-independent way.  In order to
extrapolate the measurements to the full phase space most analyses
have relied on a Monte-Carlo technique based on quasi-deuteron
kinematics \cite{Dan88,Gre91}.
This procedure hampers the comparison of the data with
more fundamental approaches.  It is to be hoped that continuous efforts
to improve on the statistics and resolution
of the measurements will make this type of
comparisons more feasible in the near future.

\begin{figure}[tbh]
\vspace{8.cm}
\caption{The fivefold
$^{16}$O($\gamma$,pn)$^{14}$N((1p1/2)$^{-1}_{\pi}$(1p1/2)$^{-1}_{\nu}$;1$^+$)
 angular cross section at E$_{\gamma}$=200~MeV, T$_p$=88~MeV and $\theta
_n$=270$^ \circ$.  The solid line involves all electric and magnetic
multipoles up to L=5, the dashed line up to L=3 and the dotted line has
the L=1 as the sole contribution.}
\label{multi}
\end{figure}

All of the {\em angular} cross sections
 presented in this work have been obtained in coplanar
kinematics. This means that the the momentum vectors of the detected
nucleons ({\bf k}$_{a}$ and {\bf k}$_{b}$) and the incoming photon beam
{\bf q}$_{\gamma}$ remain in one fixed plane.
  For all cross sections, we have averaged over
the initial photon polarization $\lambda$ and summed over the spin projections
m$_{s}$ and m$_{s'}$ of the escaping nucleons.  Unless otherwise
stated a cut-off mass $\Lambda _{\pi}$ of 1200 MeV was used. The
mean-field quantities like the bound state wave functions, the partial waves
and phase shifts for the escaping particle are selfconsistently
determined through a Hartree-Fock calculation with an effective
interaction of the Skyrme type (SkE2) \cite{Waro}.  The calculated cross
sections were observed to be rather insensitive to the choice of the
single-particle quantities.

The Feynman amplitude of eq. (17) involves a  multipole expansion over
the electric and magnetic transition operators in combination with a
partial-wave expansion for the distorted outgoing nucleon waves.    All
of these expansions were found to converge rather rapidly thus keeping
the calculations within reasonable ranges of computer time consumption.
At $E_{\gamma}$=200 MeV the expansion over the electric and magnetic
transition operators was found to converge at J=5.  This has been
illustrated in fig. \ref{multi} where we have plotted the fivefold
differential cross section $d^5 \sigma / d\Omega _p d\Omega _n dk_p$ in
planar kinematics for the
$^{16}$O($\gamma$,pn)$^{14}$N((1p1/2)$^{-1}_{\pi}$(1p1/2)$^{-1}_{\nu}$;1$^+$)
process. We have assumed that all $2h$ strength is concentrated in the
1$^+$ ground state in which the residual nucleus $^{14}$N is left.
 For the results of
fig.\ref{multi} the photon energy, the proton energy and the neutron
angle $\theta _n$ were kept fixed, so that the only remaining functional
dependence is the proton angle.  It is apparent that both the shape and
the magnitude of the cross section are determined by the lower
multipolarities, thus ensuring a fast convergence of the multipole
expansion performed for the electromagnetic interaction hamiltonian.
Regarding the
escaping particle wave functions, partial waves of higher order than
$i\frac{13}{2}$ were found not to produce any visible change in the
angular cross sections.

As a direct application of the framework presented in the previous
section we have addressed the question of how important the
distortion effects due to the interaction of the outgoing nucleons
with the residual (A-2) nucleons are.  In order to study this effect we
have performed calculations in which a plane wave description for the
detected nucleons is adopted.  These results have been compared with the
cross sections obtained after performing a full distorted wave calculation.
A plane wave description for the ejected particles can be accomplished
in the scheme presented in sect. 2.  It suffices to replace the
single-particle states p and p$'$ in the wave function of eq. (9) by
spherical Bessel functions to retain an asymptotic behaviour which in
the conventions of sect. 2.2 reads~:
\begin{eqnarray}
\left<{\bf r}_1 \mbox{\boldmath$\sigma$}_1, {\bf r}_2
\mbox{\boldmath$\sigma$}_2
    \mid \Psi_f
\right> & & \stackrel {r_1,r_2 \gg r_A} {\longrightarrow} \frac {1}
{\sqrt{A(A-1)}} \nonumber \\
 & & \times {\cal A}_{2(A-2)} \left[ \chi_{\frac{1}{2}m_s}
(\mbox{\boldmath$\sig
   ma$}_1)
e^{i{\bf k}_a {\bf \cdot r}_1}
\chi_{\frac{1}{2}m_{s'}} (\mbox{\boldmath$\sigma$}_2) e^{i{\bf k}_b {\bf \cdot
r}_2} \mid
(hh')^{-1} J_R M_R > \right]\;.
\end{eqnarray}
It is worth mentioning that when adopting a plane-wave description for
the escaping particles, the orthogonality condition between the bound
and the scattering states is lost and spurious contributions can enter
the cross sections \cite{Giutr}.  When comparing the plane wave with
the full distorted wave calculations we are actually checking in how far
plane waves are a good description for the escaping particle wave functions.
 The degree of deviation between both approaches will give us a
handle on the influence of the distortions in ($\gamma$,NN) processes.
We remind that the plane wave approximation is one of the basic
assumptions of the factorized approach to ($\gamma$,NN) processes.
In comparison with an unfactorized model, the
factorized approach extremely facilitates the interpretation of the data
and has been observed to give a fair {\em qualitative} account for the
recent $^{12}$C($\gamma$,pn) \cite{Dan88}
and $^{16}$O($\gamma$,pn) \cite{Gre91} data.

\begin{figure}
\vspace{13. cm}
\caption{The $^{12}$C($\gamma$,pn) and $^{12}$C($\gamma$,pp) strength
for knockout from the 1s shell versus photon energy.  The dashed line is
a plane wave and the solid line a full distorted wave result.}
\label{pnppfsi}
\end{figure}

In fig. \ref{pnppfsi} the calculated $^{12}$C($\gamma$,pp) and
$^{12}$C($\gamma$,pn) strength for photoinduced two-nucleon knockout
from the 1s shell is plotted versus the photon energy.  These cross
sections have been calculated with the expression (\ref{eq:inte}).  In
the process of calculating the photoabsorption strength we have summed
over all angular momentum states $ \mid J_R M_R >$ of the final nucleus. In
comparison with a plane wave calculation, the distortions are noticed to
yield a reduction of the total strength.  Apparently, the reducing
effect is stronger in the pn than in the pp channel.  As will become
clear in the course of this section, this is attributed to the
distortions affecting more  the pionic  than the isobar contributions.


\begin{figure}[tbh]
\vspace{12.4 cm}
\caption{The fivefold
$^{16}$O($\gamma$,pn)$^{14}$N((1s1/2)$^{-1}_{\pi}$(1s1/2)$^{-1}_{\nu}$)
 angular cross section at E$_{\gamma}$=200~MeV and T$_p$=48~MeV.
For the right column a full distorted wave calculation for the outgoing
pn pair was performed, whereas the left column results were obtained in
the plane wave approximation.  The upper cross sections refer to the
seagull term, the middle to the pion-in-flight and the bottom ones to a
coherent sum of both.}
\label{five1}
\end{figure}

\begin{figure}[tbh]
\vspace{12. cm}
\caption{The fivefold
$^{16}$O($\gamma$,pn)$^{14}$N((1s1/2)$^{-1}_{\pi}$(1s1/2)$^{-1}_{\nu}$)
 angular cross section in the
kinematical conditions of fig. \protect{\ref{five1}}. The
upper cross sections involve only the isobar diagrams ; the bottom ones
involve all pionic and isobar absorption diagrams.}
\label{five2}
\end{figure}

The effect of the distortions on the angular cross sections
has been investigated in  figs. \ref{five1} and \ref{five2}.  A striking
feature of all of the displayed angular cross sections is the dominance
of the back-to-back emission~: the strength is concentrated in the
kinematical range 150$^\circ \leq \mid \theta _p - \theta_n \mid \leq
210 ^\circ$.  It emerges that the distortions do not affect this picture
and that their main effect on the angular cross sections is a mere
reduction.
The angular cross sections related to the $\pi$ absorption are displayed
in fig. \ref{five1}.  Here, the distortions reduce the strength with
about 50~\%.  Remark further the very strong destructive interference
between the seagull and the pion-in-flight diagrams, affecting the
magnitude as well as the shape of the angular cross sections.  The
effect of including isobar effects is studied in fig. \ref{five2}.
Regarding the distortions, similar trends are noticed as for the pionic
terms, but the reducing effect is obviously smaller now.

\begin{figure}[tbh]
\vspace{13. cm}
\caption{The calculated photoabsorption strength for $^{12}$C($\gamma$,pp)
and $^{12}$C($\gamma$,pn) versus the photon energy.  The contributions
for knockout to the different {\em 2h} states are shown : (1s)$^{-2}$
dotted line ; (1s)$^{-1}$(1p)$^{-1}$ dashed line ; (1p)$^{-2}$
dot dashed line ; solid line : total cross section.
All absorption diagrams of fig. 3 are accounted for.}
\label{ic12pn}
\end{figure}

The total photoabsorption strength for $^{12}$C in the ($\gamma$,pp)
and the ($\gamma$,pn) channel is shown in fig. \ref{ic12pn}.  Also shown
is the contribution for pair emission from the different shells, viz.
(1s)$^{-2}$, (1s)$^{-1}$(1p)$^{-1}$ and (1p)$^{-2}$.   Apparently, the
relative strengths of the different {\em 2h} states in the ($\gamma$,pn)
channel does exhibit some
photon energy dependence.  From fig. \ref{ic12pn} it is obvious that
below the pion production threshold, the $^{12}$C($\gamma$,pn)
strength is dominated by (1p)$^{-2}$ and (1s)$^{-1}$(1p)$^{-1}$
knockout, whereas with increasing
photon energies the main contribution is coming from (1s)$^{-1}$(1p)$^{-1}$
knockout. Remark further that the calculations do not seem to follow the
naive picture in which the relative strength in the different {\em 2h}
channels is determined by the number of pn "quasi-deuteron" pairs.  In
this naive model we would expect the (1s)$^{-2}$ to
(1s)$^{-1}$(1p)$^{-1}$ ratio to be 4/16, whereas the calculations
predict that a relatively larger fraction of the ($\gamma$,pn) strength
goes through the
$^{12}$C($\gamma$,pn)$^{10}$B((1s1/2)$^{-2}$) channel.  In the pp
channel little energy dependence in the ratios between the different
{\em 2h} states channels is observed.  This has to be attributed to the
fact that in the presented calculations only one absorption mechanism is
contributing to the pp channel.  The observed energy dependencies in the
pn channel reflect the sensitivity of the different absorption mechanisms
to the shell-model structure of the final (A-2) state.

\begin{figure}
\vspace{7. cm}
\caption{Ratio of $^{12}$C($\gamma$,pn)/$^{12}$C($\gamma$,pp) absorption
strength versus the photon energy.  Dotted line : (1s)$^{-2}$ knockout ;
dashed line : (1s)$^{-1}$(1p)$^{-1}$ ; dot-dashed line : (1p)$^{-2}$ ;
solid line : ratio of the total ($\gamma$,pn) to ($\gamma$,pp) cross section.
The data are from ref. \protect{\cite{Gre91}} (circle) and ref.
\protect{\cite{Kana87}} (triangle).}
\label{ratio}
\end{figure}

Figure \ref{ratio} shows the photon energy dependence of the
calculated ratio of the total
$^{12}$C($\gamma$,pn) to the $^{12}$C($\gamma$,pp) strength.  The
results clearly illustrate that the ($\gamma$,pn)/($\gamma$,pp) ratio is
both photon energy and shell structure dependent.
Depending on the $2h$ structure of the state in which the residual (A-2)
nucleus is fed, the ratio can differ significantly from the total
photoabsorption value which is determined by a weighted average of all
of the different $2h$ contributions.  Particularly (1s)$^{-2}$ knockout
is noticed to exhibit a different overall behaviour in
($\gamma$,pn)/($\gamma$,pp) ratio.
Comparing the calculations with
the experimental estimates we observe a general overestimation of the
ratio.  This is not surprising given that we assumed a direct knockout
mechanism.  At lower photon energies, the direct ($\gamma$,pn) strength
is two orders of magnitude larger than the direct ($\gamma$,pp)
strength.  The slightest coupling between both channels, through (n,p)
rescattering,  could yield
additional strength in the pp channel and dramatically reduce the
($\gamma$,pn)/($\gamma$,pp) ratio.  In spite of the fact that for
E$_{\gamma} \geq$~250~MeV intermediate isobar creation is the most
dominant contribution to both channels, the calculated ratio for direct
knockout seems to converge around 14.  On the other hand, the data seem
to scatter around 10 in the $\bigtriangleup$(1232) resonance region.  It
remains to be investigated whether this deviation is a mere reflection
of the coupling between both channels or points towards SRC effects
putting additional strength in the ($\gamma$,pp) and ($\gamma$,pn)
channel.  The SRC could be expected to decrease the ratio of the
strongest to the weakest channel.  Other effects which deserve
to be investigated in the resonance region are off-shell effects of
$\bigtriangleup$ propagation, like summarized in the $\bigtriangleup$-hole
model \cite{delhole}, and diagrams related to the $\rho$ meson decay of the
intermediate isobar.

\begin{figure}[tbh]
\vspace{8. cm}
\caption{The $^{16}$O($\gamma$,pn) strength versus the photon energy.  The
solid
    line
is the result of the present calculation.   The data are from ref.
\protect{\cite{Carlos}} (squares ($\gamma$,1n...)), ref.
\protect{\cite{Bonnpn}} (triangle ($\gamma$,pn)) and ref.
\protect{\cite{Gre91}} (circle ($\gamma$,pn)).}
\label{io16pn}
\end{figure}

In fig. \ref{io16pn} we compare the calculated $^{16}$O($\gamma$,pn)
photoabsorption strength with the ($\gamma$,1n...) data of ref.
\cite{Carlos} and the ($\gamma$,pn) of refs. \cite{Gre91,Bonnpn}.  The
energy range of comparison is generally considered as the quasi-deuteron
region in which the photoabsorption strength is believed to be
dominated by the ($\gamma$,pn) channel.  Despite the large error bars
our absolute and parameter-free calculations seem to confirm this
picture since we exhaust a considerable fraction of the ($\gamma$,1n...)
strength and give a fair account of the ($\gamma$,pn) data points.  In
passing, it seems that there is very little room for a strong absorptive
effect in the final state interaction.
A similar but
much more pronounced effect has been found for the ($\gamma$,p) and
($\gamma$,n) channel for which direct-knockout calculations (DKO) in an
optical-model approach severely
underestimate the data \cite{gerard}.  Part of this discrepancy has been
attribu
   ted to
rescattering processes of the long-range type in the final state
interaction, which yield a substantial amount of additional strength
\cite{Gar81,Ryc88}.
It remains to be investigated how big these effects are in the
two-nucleon emission channel.

\begin{figure}[tbh]
\vspace{10. cm}
\caption{The fivefold
$^{12}$C($\gamma$,pn)$^{10}$B((1p3/2)$^{-1}_{\pi}$(1s1/2)$^{-1}_{\nu}$)
 angular cross section at E$_{\gamma}$=200~MeV, T$_p$=78~MeV and
 $\theta _n$=270$^{\circ}$ for two values of the pion cut-off mass
 ($\Lambda _{\pi}$=1200 MeV/c$^2$ and  $\Lambda _{\pi}$=800 MeV/c$^2$).
The dashed line gives the cross section for $\bf{J} ^{(\pi sea)}$,
the dotted for $\bf{J}^{(\pi sea)}$ + $\bf{J} ^{(\pi fli)}$ and the
solid for $\bf{J}^{(\pi sea)}$+$\bf{J}^{(\pi fli)}$+$\bf{J}^{(\pi
\bigtriangleup)}$.  In the insert the function F$_{1p1s}$(P=$\mid \bf{k}_p
+ \bf{k}_n - \bf{q}_{\gamma} \mid$) is plotted
versus the proton angle.}
\label{1s1p}
\end{figure}

{}From fig. \ref{ic12pn} it is obvious that the $^{12}$C($\gamma$,pn)
channel is dominated by (1s)$^{-1}$(1p)$^{-1}$ knockout at higher photon
energies. In fig. \ref{1s1p} we show the fivefold angular cross section
for the
$^{12}$C($\gamma$,pn)$^{10}$B((1p3/2)$^{-1}_{\pi}$(1s1/2)$^{-1}_{\nu}$)
reaction at E$_{\gamma}$=200 MeV, T$_p$=78 MeV and $\theta _n$=270$^{\circ}$.
The average excitation energy of the {\em 2h} state state
(1p3/2)$^{-1}_{\pi}$(1s1/2)$^{-1}_{\nu}$ was chosen to be 22 MeV.
Consequently, the neutron kinetic energy T$_n$ equals almost the
proton kinetic energy. For the
results of fig. \ref{1s1p} we have summed over the two angular momentum
states $J_R=1^-,2^-$ of the residual nucleus.
  The angular cross section is plotted for
different combinations of the photoabsorption expansion and two values
of the pion cut-off mass $\Lambda _{\pi}$.  Apparently, the shape of the
angular cross section of the type displayed in fig. \ref{1s1p}
is roughly independent of the absorption mechanism and the
value of the cut-off mass.  On the contrary, the absolute value turns
out to be sensitive to either of those.  Some theoretical ambiguities
exist with respect to the value of $\Lambda _{\pi}$.  Obviously, the
cut-off mass simply affects the magnitude of the cross section.  This
dependence illustrates the sensitivity of the ($\gamma$,NN) cross
sections to the short-range effects.  In the calculations of refs.
\cite{Giu92} the calculated ($\gamma$,pp) and (e,e$'$pp) cross
sections were observed to be very sensitive to the choice of the
short-range
Jastrow correlation functions which was introduced
as a correction to the nuclear Slater determinants.
It should be stressed, however, that in the calculation of ref.
\cite{Giu92} no hadronic form factors were introduced.  Therefore, the
Jastrow correlation function acts as the sole regularization of the $\pi$NN
vertices at short internucleon distances.
Combining hadronic form factors with a Jastrow like correlation function
to correct for the short-range behaviour in nuclear Slater determinants,
could be expected to reduce
the sensitivity of the calculated cross sections to
the Jastrow correlation function.

In an unfactorized model the angular cross section for knockout from the
single-particle orbits h and h$'$ is given by \cite{Boa89,Ryc92,Mac93} :
\begin{equation}
\frac{d^5 \sigma^{lab}} {d\Omega_p d\Omega_n dk_p} = K S_{fi}
F_{hh'}(P=\mid \bf{k}_p
+ \bf{k}_n - \bf{q}_{\gamma} \mid)  \;,
\end{equation}
where K is a kinematical factor, S$_{fi}$ determines the physics of the
photoabsorption and F$_{hh'}$ is proportional to the probability to
find a nucleon pair with zero separation and total momentum P in the
single-particle states h and h$'$.  In the insert of fig. \ref{1s1p} we
show the function F$_{1p_{3/2}1s_{1/2}}$(P) for the kinematical
conditions of the displayed cross sections.  It turns out that even in
our fully unfactorized model, including FSI effects,
 the shape of the cross section is
to a large extent determined by the probability F(P).  The absorption diagrams
s
   eem to
merely modulate the main trends set by the probability F(P).  This could
be interpreted as the theoretical justification for the observation that
the scaling mechanism of the ($\gamma$,NN) data in terms of the variable
P seems to work reasonably well.  Therefore, with the eye on future
comparisons with the data we would suggest that {\em absolute} cross sections
ar
   e
compared, because any model with a two-body type of absorption mechanism
will more or less fit the P dependence of the data, the main trends being set
by the function F(P).  As an alternative to the proposed type of
comparions on an absolute
scale, we suggest to compare angular cross sections at a fixed value of
the missing momentum P.  In this way the F(P) dependence can be ruled
out and a much larger sensitivity of the calculated cross sections to the
absorption mechanism is reached.

In contrast to what is noticed in ($\gamma$,p) and (e,e$'$p) reactions the
distorted waves do not yield considerable shifts in the strength
distributions which would be expected within a plane wave description.
A possible explanation starts from considering that the A-2
system will affect the momentum of both escaping nucleons independently.
In ($\gamma$,p) and (e,e$'$p) processes, the effect of the distortions
can to a large extent be parametrized
by introducing an effective momentum {\bf p}$_{N}^{eff}$
with which the nucleon initially escapes from the target nucleus to
arrive at the detectors with momentum {\bf p}$_{N}$ \cite{Fin77}.  The
FSI related shifts in the ($\gamma$,p) cross sections can then be
compensated for by plotting the cross sections against the {\em
effective} missing momentum {\bf p}$_{m}^{eff}$={\bf p}$_{N}^{eff}$-{\bf
q}$_{\gamma}$.
Given the back-to-back
situation, the concept of the
effective momentum could be expected to apply to each of the escaping
nucleon momenta, but the net effect on the total missing momentum ${\bf P}=
{\bf k}_p + {\bf k}_n - {\bf q}_{\gamma}$ will largely cancel out
({\bf P}$^{eff} \approx $ {\bf P}), which
might explain the relatively
small shifts induced by the distortions in two-nucleon
emission processes.

\section{Conclusion and prospects}
In conclusion, we have developed a model for the description of
electromagnetically induced
two-nucleon emission processes from finite nuclei
in a general shell-model framework.  The method relies on a partial wave
expansion for the escaping nucleon waves and a multipole expansion for
the coupling of the (virtual) photon field with the nucleus.  The basic
assumptions of the model are that the nucleon pairs escape from the
target in a one-step reaction mechanism following the absorption of the
photon on the two-body currents induced by the nucleon-nucleon
correlations in the target system.  The currents accounted for in
this work are related to one-pion exchange and
intermediate $\bigtriangleup$ isobar creation.

Within this framework we have calculated angular cross sections for the
($\gamma$,pn) and the ($\gamma$,pp) reaction off the target nuclei
$^{12}$C and $^{16}$O.   For these
purposes the coupling of the photon field to the two-body currents has
been restricted to all diagrams involving at most one intermediate
$\bigtriangleup$ line and one pion.  It has been
demonstrated that the distortions in the outgoing nucleon waves bring
about relatively small corrections to the cross sections which would be
obtained in a plane-wave model.  The main effect is that the distorted
waves yield a reduction of the cross section.  The reduction factor is
dependent on the choice of the primary photoabsorption diagrams and
the photon energy but was found not to be larger than two.
A striking feature emerging from our investigations
is that despite of the fact that the distortions can only be treated in
an unfactorized model, the distortions do not seem to affect drastically
the shape of the factorized cross section.  Consequently, the
scaling of the factorized approach in terms of the missing momentum P
seems to be somehow preserved, even when the plane wave approximation is
dropped.

In order to regularize the nucleon-nucleon correlations at short
distances, hadronic form factors had to be introduced in the calculations.
Some theoretical ambiguities exist with respect to the value of the
cut-off parameter.  The
sensitivity of the calculations to this theoretical ambiguity is
a 50~\% uncertainty
in the absolute value of the calculated cross sections.

Since angular cross sections for the ($\gamma$,NN) reaction from finite
nuclei are as yet not available, we have compared our calculations with
total ($\gamma$,1n...) and ($\gamma$,pn) photoabsorption data.
The model was found to give a fair
account of the data, pointing towards the realistic character of the adopted
model assumptions.
The ratio of the ($\gamma$,pn) strength into the different $2h$ channels
emerged to be photon energy dependent, reflecting the sensitivity of the
calculated cross sections to the nuclear structure of the final state.
A similar type of dependence was noticed for the
($\gamma$,pn)/($\gamma$,pp) ratio.\\
\vspace{2. cm}


{\Large Acknowledgement}\\
This work has been supported by the National Fund for Scientific
Research (NFWO), the Institute for Scientific Research in Industry and
Agriculture (IWONL) and in part by the NATO through the research grant
NATO-CRG920171.
\newpage
\appendix
\section{Two-body matrix elements}
We report here the reduced matrix elements of the electric and
magnetic transition operator for the two-body currents {\bf J}$^{(\pi sea)}$,
{\bf J}$^{(\pi fli)}$ and {\bf J}$^{(\pi \bigtriangleup)}$.
The transition operators are defined
according to :
\begin{eqnarray}
T_{JM}^{el}(q) & = & \sum_{M_1,M_2,\kappa =\pm 1}
\frac{i(-1)^{\delta_{\kappa,+1}}} {\hat{J}} \int d{\bf r}
 \sqrt{J+\delta _{\kappa,-1}} < J+\kappa \; M_1 \; 1 \; M_2 | J \;M >
\nonumber \\
 & & \qquad \qquad \qquad \times Y_{J+\kappa
M_1} (\Omega) j_{J+\kappa}(qr) J_{M_{2}}({\bf r}) \nonumber \\
T_{JM}^{mag}(q) & = & \sum_{M_1,M_2}
\int d{\bf r}
< J \; M_1 \; 1 \; M_2 | J \; M > Y_{J
M_1} (\Omega) j_{J}(qr) J_{M_{2}}({\bf r}) \;,
\end{eqnarray}
where {\bf J}({\bf r}) is the nuclear current operator and $\hat{J} \equiv
\sqrt{2J+1}$.
Introducing the operator
\begin{equation}
O_{JM}^{\kappa}(q)= \sum_{M_1,M_2}
\int d{\bf r}
 < J+\kappa \; M_1 \; 1 \; M_2 | J \; M > Y_{J+\kappa
M_1} (\Omega) j_{J+\kappa}(qr) J_{M_{2}}({\bf r})\;,
\end{equation}
they can be rewritten as :
\begin{eqnarray}
T_{JM}^{mag}(q) & = &  O_{JM}^{\kappa =0} \\
T_{JM}^{el}(q)  & = & \sum_{\kappa =\pm 1}
\frac{i(-1)^{\delta_{\kappa,+1}}} {\hat{J}} \sqrt{J+\delta_{\kappa,-1}}
O_{JM}^{\kappa}\;.
\end{eqnarray}
Accordingly, it suffices to calculate the matrix elements
of the
operator $O_{JM}^{\kappa}$ to determine both the electric and the
magnetic strength.   In calculating the matrix elements we
have adopted the isospin convention : $\tau _z |\pi> = + 1 |\pi>$ and
$\tau _z |\nu> = -1 |\nu>$.
For all of the expressions of the matrix elements
$<ab ;J_1 \|  O_{J}^{\kappa}(q) \|cd ; J_2 >_{as}$ in this appendix
we have assumed that the single
particle states a and c (b and
d) are of the same nature.
After lenghty but straightforward
calculations the matrix elements
for the pionic currents {\bf J}$^{(\pi sea)}$ and {\bf J}$^{(\pi fli)}$
 read :
\begin{eqnarray}
 & & <ab ;J_1 \| O_{J}^{\kappa (\pi sea)}(q) \|cd ; J_2 >_{as}   =     e
\left(\frac{f_{\pi NN}}{m_{\pi}}\right)
^{2} \frac{2}{i \pi^{3/2}} \sum_{\eta=\pm 1}
\sum_{l_4,J_3,J_4} \hat{J} \hat{J_3} \hat{J_4}
\hat{J_1} \hat{J_2} \sqrt{l_4 + \delta_{\eta ,
+ 1}}  \nonumber \\
& & \times \left(\delta_{ac,\nu}\delta_{bd,\pi}
- \delta_{ac,\pi}\delta_{bd,\nu}  \right) \nonumber \\
& & \times   \left\{ \matrix{ l_4\ & J_3 & J+\kappa \nonumber \cr
                              1    & J   & J_4 \nonumber \cr }
\right\}
   \int dp \frac{p^3}{p^2+m_{\pi}^{2}}
\left(\frac{\Lambda_{\pi}^2 - m_{\pi}^2}
{\Lambda_{\pi}^2 +p^2} \right)^{2}
 \int dr_1 \int dr_2
\nonumber \\
& & \times  < l_4 \; 0 \; J_3 \; 0 | J+\kappa \;0> (-1)^{j_c+j_d+J_2}
 \nonumber \\
& &
\times [X(j_a,j_b,J_1;j_d,j_c,J_2;l_4,J_4,J) B(a,d,l_4+\eta,l_4+\eta,l_4,pr_1)
\nonumber \\
 & & \qquad \qquad \qquad \times  B(b,c,l_4,J_3,J_4,pr_2) j_{J+\kappa}(qr_2)
\nonumber \\  & & -   (-1)^{l_4+J_4+J} X(j_a,j_b,J_1;j_d,j_c,J_2;J_4,l_4,J)
B(a,d,l_4,J_3,J_4,pr_1) \nonumber \\
& & \qquad \qquad \qquad \times   B(b,c,l_4+\eta,l_4 + \eta,l_4,pr_2)
j_{J+\kapp
   a}(qr_1)]
\end{eqnarray}

\begin{eqnarray}
& & <ab ;J_1 \|  O_{J}^{\kappa (\pi fli)}(q) \|cd ; J_2 >_{as}   =     e
\left(\frac{f_{\pi NN}}{m_{\pi}}\right)
^{2} \frac{4}{i \pi^{5/2}} \sum_{\eta,\eta',\epsilon =\pm 1}
\sum_{l_4,l'_4,J_3} \hat{J} \hat{J_3} \hat{l'_4}
\hat{J_1} \hat{J_2} \qquad \qquad
\nonumber \\ & & \times \sqrt{l_4 + \delta_{\eta ,
+ 1}}  \sqrt{l'_4 + \delta_{\eta' ,
+ 1}} \sqrt{l'_4 + \eta'+ \delta_{\epsilon ,
+ 1}} \left(\delta_{ac,\nu}\delta_{bd,\pi}
- \delta_{ac,\pi}\delta_{bd,\nu}  \right) \nonumber \\
& & \times   \left\{ \matrix{ l_4 & l'_4 & J+\kappa \nonumber \cr
                                1 &  J   & J_3 \nonumber } \right\}
\left\{ \matrix{ 1 & l'_4 & l'_4+\eta ' \nonumber \cr
                 1 & l'_4+\eta ' + \epsilon & J_3 \nonumber \cr }
 \right\}
\nonumber \\
& &  \times \int dp_1 \frac{p_1^3}{p_1^2+m_{\pi}^{2}}
\left(\frac{\Lambda_{\pi}^2 - m_{\pi}^2}
{\Lambda_{\pi}^2 +p_1^2} \right)
   \int dp_2 \frac{p_2^4}{p_2^2+m_{\pi}^{2}}
\left(\frac{\Lambda_{\pi}^2 - m_{\pi}^2}
{\Lambda_{\pi}^2 +p_2^2} \right) \nonumber \\
 & & \times \int dr_1 \int dr_2 \int dr r^2 j_{l_{4}}(p_1r)j_{l'_{4}}(p_2r)
j_{J+\kappa}(qr)
\nonumber \\
 & & \times  < l_4 \; 0 \; l'_4 \; 0 | J+\kappa \;0> (-1)^{j_c+j_d+J_2}
\times \nonumber \\
 & &
 [X(j_a,j_b,J_1;j_d,j_c,J_2;l_4,J_3,J) B(a,d,l_4+\eta,l_4+\eta,l_4,p_1r_1)
\nonumber \\
  & & \qquad \times  B(b,c,l'_4+\eta'+\epsilon,l'_4+\eta'+\epsilon,J_3,p_2r_2)
\nonumber \\  & & -  (-1)^{l_4+J_3+J} X(j_a,j_b,J_1;j_d,j_c,J_2;J_3,l_4,J)
B(a,d,l'_4+\eta '+\epsilon,l'_4+\eta '+\epsilon,J_3,p_2r_1) \nonumber \\
 & & \qquad \times   B(b,c,l_4+\eta,l_4 + \eta,l_4,p_1r_2) ]\;,
\end{eqnarray}
where,
\begin{eqnarray}
B(a,b,l,J_1,J_2,pr)  & \equiv &  <a \| j_l(pr) \left[Y_{J_{1}}(\Omega) \otimes
\sigma \right]_{J_{2}} \| b >_{r} \nonumber \\
& =  &\hat{j_a} \hat{J_2} \hat{j_b} X(l_a,1/2,j_a;l_b,1/2,j_b;J_1,1,J_2)
\hat{l_a} \hat{J_1} \sqrt{\frac{3}{2 \pi}} \nonumber \\
 & & \times (-1)^{J_1}< l_a \; 0 \; J_1 \; 0 | l_b \; 0> r^2
\varphi _a (r) \varphi _b (r)j_l(pr)\;,
\end{eqnarray}
and X denotes the 9j coefficient in
the conventions of ref. \cite{talmi} :
\begin{eqnarray}
 X(j_a,j_b,J_1;j_c,j_d,J_2;j_e,j_f,J) = \left\{ \begin{array}{lll}
    j_a \; j_b \; J_1  \nonumber \\
    j_c \; j_d \; J_2  \nonumber \\
    j_e \; j_f \; J
    \end{array} \right\} \;.
\end{eqnarray}
The matrix element for {\bf J}$^{(\pi \bigtriangleup)}$ reads :
\begin{eqnarray}
& & <ab ;J_1 \|  O_{J}^{\kappa (\pi \bigtriangleup)}(q) \|cd ; J_2 >_{as}   =
\frac{f_{\pi NN} f_{\pi N \bigtriangleup} f_{\gamma N \bigtriangleup}}
{m_{\pi}^{3}(M_{\bigtriangleup}-M_N)}
 \frac{2q}{i \pi^{3/2}} \sum_{\eta,\epsilon,\rho =\pm 1}
\sum_{l_4,J_3,J_4} \hat{J}
\hat{J_1} \hat{J_2} \qquad \qquad \qquad
\nonumber \\
& &  \int dp \frac{p^4}{p^2+m_{\pi}^{2}}
\left(\frac{\Lambda_{\pi}^2 - m_{\pi}^2}
{\Lambda_{\pi}^2 +p^2} \right)^2
 \int dr_1 \int dr_2
\nonumber \\
 & &
\times \{ (-1)^{j_c+j_d+J_2} \left(\delta_{ac,\pi}\delta_{bd,\nu}
- \delta_{ac,\nu}
\delta_{bd,\pi}  \right) \nonumber \\
& & \sum_{J_5} \frac{12}{9} R1(l_4,J_3,J_4,J_5,J,\kappa,\rho) \hat{l_4}
\hat{J_5} (-1)^{J_5+J_3}
R2(l,J_3,J_4,J,\kappa,\rho,\eta,\epsilon)
\times \nonumber \\
 & &
 [X(j_a,j_b,J_1;j_d,j_c,J_2;J_5,J_3,J) B(a,d,l_4,J_4,J_5,pr_1)
\nonumber \\
  & & \qquad \times  B(b,c,l_4+\eta+\epsilon,l_4+\eta+\epsilon,J_3,pr_2)
  j_{J+\kappa+\rho}(qr_1)
\nonumber \\  & & - (-1)^{J_3+J_5+J}  X(j_a,j_b,J_1;j_d,j_c,J_2;J_3,J_5,J)
B(a,d,l_4+\eta+\epsilon,l_4+\eta+\epsilon,J_3,pr_1) \nonumber \\
 & & \qquad \times   B(b,c,l_4,J_4,J_5,pr_2) j_{J+\kappa+\rho}(qr_2)]
\nonumber \\
& & + \frac{i 8 \sqrt{6}}{9}
\left\{ \matrix{ J_4 & J+\kappa+\rho & l_4 \nonumber \cr
                 1 & J_3 & J \nonumber \cr } \right\} (-1)^{J+\kappa}
R2(l_4,J_3,J_4,J,\kappa,\rho,\eta,\epsilon)
\nonumber \\
& &
\qquad \times [\left(\delta_{bd,\pi}-\delta_{bd,\nu}\right)
X(j_a,j_b,J_1;j_c,j_d,J_2;J_4,J_3,J)
\nonumber \\
  & & \qquad \qquad \times  C(a,c,l_4,J_4,pr_1)
B(b,d,l_4+\eta+\epsilon,l_4+\eta+\epsilon,J_3,pr_2)
  j_{J+\kappa+\rho}(qr_1)
\nonumber \\  & & \qquad +(-1)^{J_3+J_4+J}
\left(\delta_{ac,\pi}-\delta_{ac,\nu}\right)
X(j_a,j_b,J_1;j_c,j_d,J_2;J_3,J_4,J)
 \nonumber \\
 & & \qquad \qquad \times   C(b,d,l_4,l_4,J_4,pr_2)
B(a,c,l_4+\eta+\epsilon,l_4+\eta+\epsilon,J_3,pr_1) j_{J+\kappa+\rho}(qr_2)

\nonumber \\
& & - (-1)^{j_c+j_d+J_4}\left(\delta_{bc,\pi}-\delta_{bc,\nu}\right)
X(j_a,j_b,J_1;j_d,j_c,J_2;J_4,J_3,J)
\nonumber \\
  & & \qquad \qquad \times  C(a,d,l_4,J_4,pr_1)
B(b,c,l_4+\eta+\epsilon,l_4+\eta+\epsilon,J_3,pr_2)
  j_{J+\kappa+\rho}(qr_1)
\nonumber \\  & & \qquad -(-1)^{j_c+j_d+J_3+J}
\left(\delta_{ad,\pi}-\delta_{ad,\nu}\right)
  X(j_a,j_b,J_1;j_d,j_c,J_2;J_3,J_4,J)
 \nonumber \\
 & & \qquad \qquad \times   C(b,c,l_4,J_4,pr_2)
B(a,d,l_4+\eta+\epsilon,l_4+\eta+\epsilon,J_3,pr_1)
 j_{J+\kappa+\rho}(qr_2)] \}\;,
\end{eqnarray}
where,
\begin{eqnarray}
R1(l,J_1,J_2,J_3,J,\kappa,\rho) & \equiv & \sum_{J_{4}} (2J_4+1)
\left\{ \matrix{J+\kappa+\rho & J_2 & l \nonumber \cr
                J_1           & 1   & J_4 \nonumber \cr } \right\}
\left\{ \matrix{J_3 & 1 & J_2 \nonumber \cr
                J_4 & J_1 & J \nonumber \cr } \right\} \nonumber \\
& & \times \left\{ \matrix{ 1 & 1 & 1 \nonumber \cr
                 J_4 & J+\kappa+\rho & J \nonumber \cr } \right\}\;,
\nonumber \\
R2(l,J_1,J_2,J,\kappa,\rho,\eta,\epsilon) & \equiv &
\sqrt{J+\kappa+\delta_{\rho , +1}}
\sqrt{l+\delta_{\eta , +1}}
\sqrt{l+\eta+\delta_{\epsilon , +1}} \nonumber \\
& & \times \sqrt{2(J+\kappa+\rho)+1} \hat{J_1} \hat{J_2}
\left\{ \matrix{ 1 & l & l+\eta \nonumber \cr
                 1 & l+\eta+\epsilon & J_1 \nonumber \cr } \right\}
\nonumber \\
& & \times \left\{ \matrix{ 1 & 1 & 1 \nonumber \cr
                 J+\kappa & J+\kappa+\rho & J \nonumber \cr } \right\}
\left( \matrix{ l & J_2 & J+\kappa +\rho \nonumber \cr
                 0 & 0 & 0 \nonumber \cr } \right) \nonumber \;, \\
C(a,b,l,J_1,pr) & \equiv & <a \| j_l(pr) Y_{J_{1}}(\Omega) \|b> _{r}
\end{eqnarray}

The integrals over the pion momenta in the above matrix elements can be
performed analytically using contour integration techniques
\cite{suzuki} \cite{ama93}.
As an example we mention the expressions for the momentum integrals in
the pion-in-flight matrix element.  It can be shown that :
\begin{eqnarray}
& & \frac{2}{\pi} \int dp \frac{p^3}{p^2+m_{\pi}^{2}}
\left(\frac{\Lambda_{\pi}^2 - m_{\pi}^2}
{\Lambda_{\pi}^2 +p^2} \right) j_l(pr_1)j_{l+\eta}(pr_2) =
 \qquad \qquad \qquad \qquad  (\eta = \pm 1) \nonumber \\
& & -im_{\pi}^2 \left[ \theta(r_1-r_2)
h_l^{(1)}(im_{\pi}r_1) j_{l+\eta}(im_{\pi}r_2)
+ \theta(r_2-r_1)
h_l^{(1)}(im_{\pi}r_2) j_{l+\eta}(im_{\pi}r_1) \right] \nonumber \\
& & +i\Lambda_{\pi}^2 \left[ \theta(r_1-r_2) h_l^{(1)}(i\Lambda_{\pi}r_1)
j_{l+\eta}(i\Lambda_{\pi}r_2)
+ \theta(r_2-r_1) h_l^{(1)}(i\Lambda_{\pi}r_2)
j_{l+\eta}(i\Lambda_{\pi}r_1) \right] \\
& & \frac{2}{\pi} \int dp \frac{p^4}{p^2+m_{\pi}^{2}}
\left(\frac{\Lambda_{\pi}^2 - m_{\pi}^2}
{\Lambda_{\pi}^2 +p^2} \right) j_l(pr_1)j_{l+\eta}(pr_2) =
 \qquad \qquad \qquad \qquad  (\eta = 0,\pm 2) \nonumber \\
& & m_{\pi}^3 \left[ \theta(r_1-r_2)
h_l^{(1)}(im_{\pi}r_1) j_{l+\eta}(im_{\pi}r_2)
+ \theta(r_2-r_1)
h_l^{(1)}(im_{\pi}r_2) j_{l+\eta}(im_{\pi}r_1) \right] \nonumber \\
& & - \Lambda_{\pi}^3 \left[ \theta(r_1-r_2) h_l^{(1)}(i\Lambda_{\pi}r_1)
j_{l+\eta}(i\Lambda_{\pi}r_2)
+ \theta(r_2-r_1) h_l^{(1)}(i\Lambda_{\pi}r_2)
j_{l+\eta}(i\Lambda_{\pi}r_1) \right] \;.
\end{eqnarray}
The momentum integrals in the other matrix elements take on similar
expressions. The radial integrals in the above matrix elements
have to be performed
numerically.


\begin{thebibliography} {99}
\bibitem{Got58}
Kurt Gottfried, Nucl. Phys. {\bf 5} (1958) 557.
\bibitem{Wei70}
W. Weise, M.G. Huber and M. Danos, Z. Phys. {\bf236} (1970) 176.
\bibitem{Bra71}
E. Bramanis, Nucl. Phys. {\bf A175} (1971) 17.
\bibitem{Wak83}
M. Wakamatsu and K. Matsumoto, Nucl. Phys. {\bf A392} (1983) 323.
\bibitem{Bof91}
S. Boffi and M.M. Giannini, Nucl. Phys. {\bf A533} (1991) 441\,.
\bibitem{Boa89}
L. Boato and M.M. Giannini, J. Phys. {\bf G15} (1989) 1605.
\bibitem{Ryc92}
J. Ryckebusch, L. Machenil, M. Vanderhaeghen and M. Waroquier, Phys.
Lett. {\bf B291} (1992) 213.
\bibitem{Gar81}
M. Gari and H. Hebach, Phys. Rep. {\bf 72} (1981) 1.
\bibitem{Ryc88}
J. Ryckebusch, M. Waroquier, K. Heyde, J. Moreau and D. Ryckbosch, Nucl. Phys.
{\bf A476} (1988) 237.
\bibitem{Bau90}
T.S. Bauer, Nucl. Phys. {\bf A546} (1992) 181c.
\bibitem{Giu92}
C. Giusti, F.D. Pacati and M. Radici, Nucl. Phys. {\bf A546} (1992) 607.
\bibitem{Mah}
C. Mahaux and H. Weidenm\"{u}ller, {\em in} A shell model approach to
nuclear reactions (North Holland, Amsterdam, 1969).
\bibitem{Mac93}
L. Machenil, M. Vanderhaeghen, J. Ryckebusch and M.
Waroquier, to be published.
\bibitem{Bonn}
R. Machleidt, K. Holinde and Ch. Elster, Phys. Rep. {\bf 149} (1987) 1.
\bibitem{Sas92}
T. Sasakawa, S. Ishikawa, Y. Wu and T-Y. Saito, Phys. Rev. Lett. {\bf
68} (1992) 3503.
\bibitem{For66}
T. De Forest and J.D. Walecka, Adv. Phys. {\bf 15} (1966) 1.
\bibitem{talmi}
I. Talmi, Simple Models of Complex Nuclei : The Shell Model and
Interacting Boson Model, (Harwood Academic Publishers, Chur, 1993).
\bibitem{Dan88}
S.N. Dancer,I.J.D. MacGregor, J.R.M. Annand, I. Anthony, G.I.
Crawford,
S.J. Hall, J.D. Kellie, J.C. McGeorge, G.J. Miller, R.O. Owens,
P.A. Wallace, D. Branford, S.V. Springham, A.C. Schotter,
B. Schoch, J. M. Vogt, R. Beck and G. Liesenfeld,
Phys. Rev. Lett. {\bf 61} (1988)1170 \,.
\bibitem{Gre91}
I.J.D. MacGregor, J.R.M. Annand, I. Anthony, S.N. Dancer, S.M. Doran,
S.J. Hall, J.D. Kellie, J.C. McGeorge, G.J. Miller, R.O. Owens,
P.A. Wallace, B. Schoch, H. Schmieden, J. M. Vogt and S. Klein,
 Nucl. Phys.  {\bf A533} (1991) 269.
\bibitem{Waro}
M. Waroquier, J. Ryckebusch, J. Moreau, K. Heyde, N. Blasi,
S.Y. van der Werf and G. Wenes, Phys. Rep. {\bf 148} (1987) 249.
\bibitem{Giutr}
C. Giusti and F.D. Pacati, Proc. of the 5th Workshop on Perspectives in
Nuclear Physics at Intermediate Energies
, Eds. S. Boffi, C. Ciofi degli Atti, M. Giannini (World
Scientific, Singapore 1992) 300.
\bibitem{Kana87}
M. Kanazawa, S. Homma, M. Koike, Y. Murata, H. Okuno, F. Soga and N.
Yoshikawa, Phys. Rev. {\bf C35} (1987) 1828.
\bibitem{delhole}
J.K. Koch and N. Ohtsuka, Nucl. Phys. {\bf A435} (1985) 765.
\bibitem{Carlos}
P. Carlos, H. Beil, R. Berg\'{e}re, B.L. Berman, A. Lepr\^{e}tre and
A. Veysi\'{e}re, Nucl. Phys. {\bf A378} (1982) 317.
\bibitem{Bonnpn}
H. Hartmann, H. Hoffmann, B. Mecking and G. N\"{o}ldeke, Proc. Int.
Conf. on Photonuclear reactions and Applications, ed. B.L. Berman
(Lawrence Livermore Laboratory, Livermore, 1973) vol. 2, 967.
\bibitem{gerard}
G. van der Steenhoven and H.P. Blok, Phys. Rev. {\bf C42} (1990) 2597.
\bibitem{Fin77}
D.J.S. Findlay and R.O. Owens, Nucl. Phys. {\bf A292} (1977) 53.
\bibitem{suzuki}
T. Suzuki, Nucl. Phys. {\bf A495} (1989) 581.
\bibitem{ama93}
J.E. Amaro, G. Co' and A.M. Lallena, Ann. Phys. {\bf 221} (1993) 306.
\end{thebibliography}
\end{document}